\documentclass[prd,showpcs,amsmath,amssymb,nofootinbib,preprintnumbers,balancelastpage,longbibliography,superscriptaddress,notitlepage]{revtex4}
\usepackage{multirow}
\usepackage{epsfig}
\usepackage{amsmath}
\usepackage{bm}
\usepackage{times}
\usepackage{graphicx}
\usepackage{color}
\usepackage{slashed}
\usepackage{graphicx}
\usepackage{amsmath}
\usepackage{tikz}
\usetikzlibrary{positioning,shapes}
\usepackage{relsize} 
\usepackage[latin1]{inputenc}
\usepackage{tikz}
\usetikzlibrary{trees}
\usetikzlibrary{decorations.pathmorphing}
\usetikzlibrary{decorations.markings}
\usetikzlibrary{positioning,arrows}
\usetikzlibrary{decorations.pathmorphing}
\usetikzlibrary{decorations.markings}
\usetikzlibrary{decorations.pathreplacing,calc}
\usetikzlibrary{decorations.pathmorphing,decorations.markings,trees,positioning,arrows}   
\newif\ifmirrorsemicircle

\def\bea{\begin{eqnarray}}
\def\eea{\end{eqnarray}}
\def\bean{\begin{equation*}}
\def\eean{\end{equation*}}

\def\bea{\begin{eqnarray}}
\def\eea{\end{eqnarray}}
\def\bean{\begin{equation*}}
\def\eean{\end{equation*}}

\begin{document}

\preprint{UCI-HEP-TR-2016-20}

\title{Asymmetric Dark Matter and Baryogenesis from $SU(2)_\ell$}

\author{Bartosz~Fornal}
\affiliation{Department of Physics and Astronomy, University of California, Irvine, CA 92697, USA}
\affiliation{Department of Physics, University of California, San Diego, 9500 Gilman Drive, La Jolla, CA 92093, USA}
\author{Yuri~Shirman}
\affiliation{Department of Physics and Astronomy, University of California, Irvine, CA 92697, USA}
\author{Tim~M.~P.~Tait}
\affiliation{Department of Physics and Astronomy, University of California, Irvine, CA 92697, USA}
\author{Jennifer~Rittenhouse~West}
\affiliation{Department of Physics and Astronomy, University of California, Irvine, CA 92697, USA}
\date{\today}

\begin{abstract}
We propose a theory in which the Standard Model gauge symmetry is extended by a new $SU(2)_\ell$ group acting nontrivially on the lepton sector which is
spontaneously broken at the TeV scale. Under this $SU(2)_\ell$ the ordinary leptons form doublets along with new lepton partner fields. This construction naturally contains a dark matter candidate, the partner of the right-handed neutrino, stabilized by a residual global $U(1)_\chi$ symmetry. We show that one can explain baryogenesis through
an asymmetric dark matter scenario, in which generation of related asymmetries in the dark matter and baryon sectors is driven by the $SU(2)_\ell$ instantons during a first order phase transition in the early universe. 
\vspace{11mm}
\end{abstract}

\maketitle

\section{Introduction}

Evidence for the existence of dark matter (DM) composing roughly $25\%$ of the matter in the Universe is extremely compelling and includes measurements of galactic rotation curves, gravitational lensing, cosmic microwave background anisotropy and $X$-ray emission from elliptical galaxies. The Standard Model (SM) of particle physics in its current form does not account for the presence of DM. However, if DM couples appreciably to visible matter, it is reasonable to expect that it represents an extension of the SM built on similar principles, i.e. those of a gauge theory. One such possibility is to postulate  a larger gauge symmetry group.  Ideally, one would like this expanded symmetry
to play a role in explaining why DM is stable on the scale of the age of the Universe, such that it does not decay into SM states.

Another problem associated with  the SM is its inability to explain the current matter-antimatter asymmetry of the Universe. Two of the Sakharov conditions \cite{Sakharov:1967dj}, the requirement of a first order phase transition and sufficient amount of $CP$ violation, are likely not fulfilled in the SM alone. There are many ideas for how to fix this by introducing new fields in the theory. Similarly to the approach used for the DM problem, one class of potential solutions relies on extensions of the Standard Model gauge group. For example, it was shown \cite{Shu:2006mm} that the breakdown of a new gauge group can overcome the SM difficulties and provide a framework for successful baryogenesis. 

It is tempting to combine these ideas, positing that the DM interactions with the SM are responsible both for generating its abundance as well as generating the observed baryon asymmetry.  This is the philosophy behind asymmetric dark matter (ADM) models \cite{Nussinov:1985xr,Kaplan:1991ah,Hooper:2004dc,Kaplan:2009ag,Petraki:2013wwa,Zurek:2013wia}, in which the DM and baryon asymmetries are intimately related.  A typical feature of these models is that the natural scale for the DM mass is $\sim  \rm GeV$, which requires light messengers in order to realize a large enough DM annihilation cross section to annihilate away the symmetric component. 
In this paper we construct an ADM model that naturally contains such particles.

As a starting point it is natural to consider gauged extensions of the SM global symmetries. The possibility of gauging lepton and baryon number was considered in \cite{Pais:1973mi, Tosa:1982pv, Rajpoot:1987yg, Foot:1989ts, Carone:1995pu, Georgi:1996ei}. Unfortunately, models based on this approach offer only a limited possibility of explaining the primordial baryon asymmetry \cite{FileviezPerez:2010gw,Duerr:2013dza,Schwaller:2013hqa}. More recently, a baryogenesis mechanism based on a non-Abelian extension of the SM baryon number and color in $SU(4)$ gauge group was proposed in \cite{Fornal:2015one,Fornal:2016boa}. Such an extension successfully unifies  DM with the SM baryons, but ultimately relies on unspecified UV physics represented by higher-dimensional operators to generate the asymmetry.

In this paper we follow the general approach of \cite{Fornal:2016boa} and extend the SM gauge group by an additional $SU(2)_{\ell}$ gauge symmetry under which the SM leptons transform nontrivially, promoting them to $SU(2)_{\ell}$ doublets along with additional partner fields.  A lepton number assignment is extended to the partner fields, thus generalizing the SM lepton number.  The $SU(2)_{\ell}$ symmetry is spontaneously broken via a vacuum expectation value (vev) of a new leptonic Higgs field $\Phi$. All new matter fields introduced in the model obtain vector-like masses after $SU(2)_\ell$ breaking. The lightest of these particles will be stable due to residual global symmetries and provides a dark matter candidate of ADM framework. During the $SU(2)_\ell$ phase transition, $SU(2)_\ell$ sphalerons generate both DM and lepton number, with the latter later converted to baryon number by electroweak sphalerons.  A similar use of sphalerons is described in the aidnogenesis scenario \cite{Blennow:2010qp}.

\section{Model}
\label{sec:model}

We extend the gauge symmetry of the Standard Model to:
\begin{eqnarray}
&&SU(3)_c \times SU(2)_{W} \times U(1)_Y  \times SU(2)_{\ell}\ \ .
\end{eqnarray}
 The SM leptons reside in upper components of $SU(2)_\ell$ doublets, while the new fermions (denoted by a tilde) reside in lower components of $SU(2)_\ell$ doublets:
\bea
&& \hspace{-10mm}\hat{l}_L \equiv \left(\!
  \begin{array}{c}
    \l_{L}\\
    \tilde{l}_{L} \\
  \end{array}\!
\right)  , \ \ \ \ \hat{e}_{R} \equiv \left(\!\!
  \begin{array}{c}
    e_R\\
    \tilde{e}_R \\
  \end{array}\!\!
\right)  , \ \ \ \ \hat{\nu}_{R} \equiv \left(\!\!
  \begin{array}{c}
    \nu_R \\
    \tilde{\nu}_R \\
  \end{array}\!\!
\right)\,.
\eea
To maintain cancellation of  $SU(2)_W\times U(1)_Y$ anomalies and allow new fermions to acquire Dirac masses after $SU(2)_\ell$ symmetry breaking we also introduce a set of leptons that are neutral under $SU(2)_\ell$,
\bea
&& {l}'_{R}  , \ \ \ \ {e}'_{L}  , \ \ \ \ {\nu}'_{L} \ .
\eea
Finally, we introduce two Higgs doublets charged under $SU(2)_\ell$. The second doublet allows us to introduce two ingredients necessary for a realistic ADM model: sufficient CP violation to catalyze the production of an adequate baryon asymmetry to match observation, and an annihilation channel allowing for a sufficient annihilation of the symmetric component of the DM.
The quantum numbers for all the relevant particles, including two $SU(2)_\ell$ Higgs doublets (discussed below) are summarized in
Table \ref{table2}. \\

{\renewcommand{\arraystretch}{1.4}\begin{table}[t!]
\begin{center}
    \begin{tabular}{| c || c | c | c |}
    \hline
       \ \ \ \ \ Field \ \ \ \ \  & \raisebox{0ex}[0pt]{$ \ \ \ SU(2)_\ell \ \ \ $} & \raisebox{0ex}[0pt]{$ \ \ \ SU(2)_W \ \ \ $} & \raisebox{0ex}[0pt]{$ \ \ \ U(1)_Y \ \ \  $}  
       \\ \hline\hline
         \ \ $\hat{l}_{L} =  \left(\!\begin{array}{c}
    l_L\\
    \tilde{l}_L \\
  \end{array} \!\right) \ \ $ & $2$ & $2$ & $-1/2$  \\ \hline
         $\hat{e}_R =  \left(\!\!\begin{array}{c}
    e_R\\
    \tilde{e}_R \\
  \end{array} \!\!\right)$ & $2$ & $1$ &  $-1$  \\ \hline
          $\hat{\nu}_R = \left(\!\!\begin{array}{c}
    \nu_R\\
    \tilde{\nu}_R \\
  \end{array} \!\!\right)$ & $2$ & $1$ &  $0$  \\ \hline
               $l'_{R}$ & $1$ & $2$ & $-1/2$  \\ \hline
         $e'_L$ & $1$ & $1$ &  $-1$  \\ \hline
          $\nu'_L$ & $1$ & $1$ &  $0$  \\ \hline
         ${\Phi}_1, \Phi_2$ & $2$ & $1$ &  $0$  \\ \hline
    \end{tabular}
\end{center}
\caption{\small{Fields and their representations under the gauge symmetries
$SU(2)_{\ell} \times SU(2)_{W} \times U(1)_Y$.}}
\label{table2}
\end{table}}

To allow for a spontaneous breaking of $SU(2)_\ell$, we introduce the following scalar potential for $SU(2)_\ell$ doublets $\Phi_i$:
\bea
V(\Phi_1,\Phi_2) &=&  m_1^2 |\Phi_1|^2 + m_2^2 |\Phi_2|^2+ (m_{12}^2 \Phi_1^\dagger \Phi_2 + \rm{h.c.}) + \lambda_1 |\Phi_1|^4 +\lambda_2 |\Phi_2|^4 +\,\lambda_3 |\Phi_1|^2  |\Phi_2|^2  \nonumber\\
 && + \lambda_4 |\Phi_1^\dagger\Phi_2|^2 +  \Big[ \tilde{\lambda}_5 \Phi_1^\dagger \Phi_2 |\Phi_1|^2 +\tilde{\lambda}_6 \Phi_1^\dagger \Phi_2 |\Phi_2|^2 + \tilde{\lambda}_7 (\Phi_1^\dagger \Phi_2)^2 + \ {\rm h.c.}\Big]
 \label{eq:V}
\eea
where we have imposed a $U(1)_1$ symmetry (discussed below) to ensure that the dark matter is ultimately stable.
In addition, while we do not explicitly show $\Phi_i$ interactions with the SM Higgs, such interactions are generically present at tree level and are induced radiatively even if not present.  This interaction plays an important role
in allowing the lightest component of $\Phi$ to decay through induced mixing with the SM Higgs, with constraints discussed in Sec.~\ref{DMsec}.

The potential contains four complex parameters: $m_{12}^2$, $\tilde{\lambda}_5$, $\tilde{\lambda}_6$, and $\tilde{\lambda}_7$. 
For generic parameters, one phase
can be rotated away by redefining the phase of the combination $\Phi_1^\dagger \Phi_2$ (the only combination appearing in
the potential), leaving three physical phase combinations \cite{Gunion:2005ja}.

It is easy to choose parameters so that $SU(2)_\ell$ is completely broken by $\Phi_{1,2}$ vevs.  The potential, Eq.~(\ref{eq:V}), is structurally identical to a two Higgs
doublet model (with the global $U(1)_1$ playing the role of the SM's gauged $U(1)_Y$ hypercharge) 
and admits a similarly rich array of mass eigenstates for the physical bosons.  The vacuum
can be parameterized by $v_\ell = \sqrt{v_1^2 + v_2^2}$ and $\tan \beta = v_1 / v_2$, where $v_1$ and $v_2$ are
the vevs of the two doublets, respectively, $\langle \Phi_1 \rangle = \frac{1}{\sqrt{2}}(0, v_1)^T$ and $\langle \Phi_2 \rangle = \frac{1}{\sqrt{2}}(0, v_2)^T$.  There
is a spectrum of five physical scalar Higgs bosons which are mixtures of the original CP-even and CP-odd components of the doublets $\Phi_1$ and $\Phi_2$.

The Yukawa interactions consistent with the gauge symmetries are given by
\bea
 \mathcal{L}_{\rm Y} &=&   \sum_i \left(Y_l^{ab}\,\bar{\hat{l}}_L^a \, \Phi_i\,  {l'}_{\!\!R}^{b}+ Y_e^{ab} \, \bar{\hat{e}}_R^a \, \Phi_i\, {e'}_{\!\!L}^{b}
   + Y_\nu^{ab}\,\bar{\hat{\nu}}_R^a \, \Phi_i\, {\nu'}_{\!\!L}^{b}\right)\nonumber\\  \nonumber\\[-7pt]
 &+&   y_e^{ab}\, \bar{\hat{l}}_L^a \,H \,\hat{e}_R^b + y_\nu^{ab}\, \bar{\hat{l}}_L^a\, \tilde{H} \,\hat{\nu}_R^b + {y'}_{\!\!e}^{ab}\, \bar{l'}_{\!\!R}^a \,H \,{e'}_{\!\!L}^b + {{y'}}_{\!\!\nu}^{ab}\, \bar{l'}_{\!\!R}^a\,\tilde{H} \, {\nu'}_{\!\!L}^b + {\rm h.c.} \ ,
  \label{lagr}
\eea
where $a$ and $b$ are flavor indices.
After electroweak symmetry breaking the Yukawa matrices $y^{ab}_e$ and $y^{ab}_\nu$ lead to the usual lepton mass matrices (with Dirac neutrino masses). 
The newly introduced Yukawa matrices $Y_l$, $Y_e$, $Y_\nu$, $y^\prime_e$ are responsible for generation of Dirac mass terms between the lepton partners and spectators after $SU(2)_\ell$ breaking.
In addition, the full mass matrix for new fields includes mixing between electroweak singlets and doublets induced by the SM Higgs vev:
\bea
& & \hspace*{-0.75cm}
\frac{1}{\sqrt{2}}
\left(
\overline{\tilde{\nu}}_{\!L} ~~ \overline{\nu}^\prime_L
\right)
 \left(\!
  \begin{array}{cc}
    Y_l \,v_\ell & y_\nu v \\
    {y'}_{\!\!\nu}^\dagger v & Y_\nu^\dagger  v_\ell \\
  \end{array}\!
\right)
\left(\!\!
  \begin{array}{c}
   \nu^\prime_R \\
    \tilde{\nu}_R \\
  \end{array}\!\!
\right)+ \frac{1}{\sqrt{2}}
\left(
\overline{\tilde{e}}_L ~~ \overline{e}^\prime_L
\right)
 \left(\!
  \begin{array}{cc}
    Y_l \,v_\ell & y_e v \\
    {y'}_{\!\!e}^\dagger v &Y_e^\dagger  v_\ell \\
  \end{array}\!
\right)
\left(\!\!
  \begin{array}{c}
   e^\prime_R \\
    \tilde{e}_R \\
  \end{array}\!\!
\right)
+ {\rm h.c.} \ ,
\eea 
where each field should be understood as a three-component vector in flavor space, with the Yukawa couplings $3 \times 3$ complex matrices, and the vevs $v$ and $v_\ell$ belong to the SM Higgs and the new Higgses, respectively.  

The couplings
$y_e$ and $y_\nu$ are required to reproduce the SM flavor structure, and are thus tiny.  If in addition $y'_{\nu,e} v \ll Y_{\ell, \nu, e} v_\ell$, there is little mixing
between doublets and singlets, and negligible contributions to precision electroweak observables.  We 
find it convenient to choose parameters in this technically natural regime.

\section{Baryogenesis}

In this section, we discuss the $SU(2)_\ell$ phase transition, which provides the out of equilibrium condition necessary to evolve a nonzero baryon number.

\subsection{Non-perturbative dynamics}

{\renewcommand{\arraystretch}{1.4}\begin{table*}[t!]
\begin{center}
    \begin{tabular}{| c || c | c | c || c | c | c || c | c | c |}
    \hline \multirow{2}{*}{} & 
    \multicolumn{9}{c|}{Symmetries}  \\ \cline{2-10} & \multicolumn{3}{c||}{Exact \& Approximate}& \multicolumn{3}{c||}{Lepton Basis}& \multicolumn{3}{c|}{Low Energy}\\
    \hline
       \ \ \ \ \ Field \ \ \ \ \  & \raisebox{0ex}[0pt]{$ \ \ U(1)_1 \ \ $} & \raisebox{0ex}[0pt]{$ \ \ U(1)_2 \ \ $} & \raisebox{0ex}[0pt]{$ \ \ U(1)^\prime \ \ $} & \raisebox{0ex}[0pt]{$ \ \ U(1)_1 \ \  $} & \raisebox{0ex}[0pt]{$ \ \ U(1)_L \ \  $} & \raisebox{0ex}[0pt]{$ \ \ U(1)_\chi \ \  $} & \raisebox{0ex}[0pt]{$ \ \ U(1)_D \ \  $} & \raisebox{0ex}[0pt]{$ \ \ U(1)_L \ \  $} & \raisebox{0ex}[0pt]{$ \ \ U(1)_\chi \ \  $}
        
       \\ \hline\hline
         \ \ $\hat{l}_{L} =  \left(\!\begin{array}{c}
    l_L\\
    \tilde{l}_L \\
  \end{array} \!\right) \ \ $ & $0$ & $1$ & $1$ & $0$  & $1$ & $0$ & $\!\begin{array}{c} -1\\ 1 \\ \end{array}$  & $1$ & $0$ \\ \hline
         $\hat{e}_R =  \left(\!\!\begin{array}{c}
    e_R\\
    \tilde{e}_R \\
  \end{array} \!\!\right)$ & $0$ & $1$ & $1$ & $0$  & $1$ & $0$ & $\!\begin{array}{c} -1\\ 1 \\ \end{array}$ & $1$ & $0$ \\ \hline
          $\hat{\nu}_R = \left(\!\!\begin{array}{c}
    \nu_R\\
    \tilde{\nu}_R \\
  \end{array} \!\!\right)$ & $0$ & $1$ & $-1$ & $0$  & $0$ & $1$ & $\!\begin{array}{c} -1\\ 1 \\ \end{array}$ & $0$ & $1$ \\ \hline
               $l^\prime_{R}$ & $1$ & $1$ & $1$ & $1$ & $1$ & $0$ & $1$ & $1$ & $0$ \\ \hline
         $e^\prime_L$ & $1$ & $1$ & $1$ & $1$ & $1$ & $0$ & $1$ & $1$ & $0$ \\ \hline
          $\nu^\prime_L$ & $1$ & $1$ & $-1$ & $1$ & $0$ & $1$ & $1$ & $0$ & $1$ \\ \hline
         $\Phi_i,~~i=1,2$ & $-1$ & $0$ & $0$ & $-1$ & $0$ & $0$  & $\!\begin{array}{c} -2\\ 0 \\ \end{array}$ & $0$ & $0$ \\ \hline
    \end{tabular}
\end{center}
\caption{\small{Charges under global $U(1)$ symmetries. The first two columns represent charges under exact global symmetries of the Lagrangian. The next four columns represent charges under exact and approximate global symmetries when $y_\nu$ and $y_\nu^\prime$ can be neglected. The last three columns represent charges under exact and approximate global symmetries in low energy physics.}}
\label{tab:global}
\end{table*}}

Non-perturbative dynamics lead to simultaneous lepto- and DM-genesis through sphaleron processes during the $SU(2)_\ell$ phase transition. To understand the generation of lepton and DM numbers during the $SU(2)_\ell$ transition we first need to analyze the global symmetries of the model. The theory posesses two anomaly-free global $U(1)$ symmetries consistent with the gauge structure and Yukawa interactions in Eq.~(\ref{lagr}) (see the first panel of Table~\ref{tab:global}).  In a realistic model, neutrino Yukawa couplings $y_\nu$ and $y_\nu^\prime$ must be small, therefore an additional approximate global symmetry $U(1)^\prime$ exists. The $U(1)^\prime$ global symmetry is anomalous under $SU(2)_\ell$ and thus will be broken by instanton-generated interactions. For our purposes it is convenient to construct linear combinations of the $U(1)_2$ and $U(1)^\prime$ symmetries that will correspond to generalized lepton and DM numbers. Charge assignments under these symmetries, $U(1)_L$ and $U(1)_\chi$ respectively, are shown in the middle panel of Table~\ref{tab:global}. 

The $U(1)_1$ symmetry is spontaneously broken by the $\Phi_1$ and $\Phi_2$ vevs.  However, a global $U(1)_D$ subgroup of $SU(2)_\ell\times U(1)_1$ survives. This unbroken $U(1)_D$ is a diagonal combination of the $U(1)_1$ and the $U(1)$ group generated by the $\tau_3$ generator of $SU(2)_\ell$. The charges of the fields in the low energy theory are shown in the last panel of Table~\ref{tab:global}. Note that the charges of light fields under $U(1)_D$ are given by the sum of lepton and DM charges and thus $U(1)_D$ is not visible in low energy physics. However, $U(1)_D$ distinguishes between the SM leptons and new particles and thus will be responsible for the stability of the DM.

Both $U(1)_L$ and $U(1)_\chi$ are individually anomalous under $SU(2)_\ell$ interactions. On the other hand, the sum of lepton and DM numbers will be conserved since it corresponds to an anomaly free $U(1)_2$ symmetry. This means that $SU(2)_\ell$ instantons generate effective interactions in low energy theories that break $U(1)_L$ and $U(1)_\chi$ individually while conserving the sum of the two charges, $L+\chi$. For illustrative purposes, it is convenient to consider a one flavor toy model. The $SU(2)_\ell$ instantons generate an effective 4-fermion interaction that involves all of the doublets of $SU(2)_{\ell}$. Applying results from \cite{Morrissey:2005uza}, the one-family sphalerons can be represented as a dimension six operator,
\bea
\mathcal{O}_{\rm eff} &\sim& \epsilon_{ij} \ \Big[(l_L^i \cdot \bar{{\nu}}_R)(l_L^j \cdot \bar{{e}}_R) - (l_L^i \cdot \bar{{\nu}}_R)(\tilde{l}_L^j \cdot \bar{\tilde{e}}_R) + \ (l_L^i \cdot \tilde{l}_L^j)(\bar{\nu}_R \cdot \bar{\tilde{e}}_R)\nonumber\\
&&\ \ \ \  - \  (l_L^i \cdot \tilde{l}_L^j)(\bar{\tilde{\nu}}_R \cdot \bar{{e}}_R) +   (\tilde{l}_L^i \cdot \bar{\tilde{\nu}}_R)(\tilde{l}_L^j \cdot \bar{\tilde{e}}_R) - (\tilde{l}_L^i \cdot \bar{\tilde{\nu}}_R)({l}_L^j \cdot \bar{{e}}_R)\Big], \ \ \ \ \ 
\label{inst}
\eea
where the dots denote Lorentz contractions, $i, j$ are $SU(2)_W$ indices and the $SU(2)_\ell$ indices have been expanded out. 

It is easy to see that instanton-induced interactions in Eq.~(\ref{inst}) violate lepton and DM numbers by $\Delta L =-1$ and $\Delta \chi = 1$ respectively. Consider, for example, the last term in Eq.~(\ref{inst}). It is responsible for processes $\nu_L \,\tilde{e}_L \rightarrow \tilde{\nu}_R \ {e}_R$ and $\tilde{\nu}_L \,e_L \rightarrow \tilde{\nu}_R \ {e}_R$. Since $l_L, \tilde{l}_L, e_R, \tilde{e}_R$ have $L=1$ while $\nu_R, \tilde{\nu}_R$ have $\chi=1$, the instanton-induced interactions violate lepton and DM numbers by $\Delta L =-1$ and $\Delta \chi = 1$.

The generalization to the three generation model is straightforward, leading to a 12 fermion operator. At zero temperature the instanton operator is exponentially suppressed,
but at high temperatures the $SU(2)_\ell$ symmetry is restored, and lepton- and DM-number violating interactions are unsuppressed. The combined effect of all instanton-induced interactions (\ref{inst}) is calculated by numerically solving the diffusion equations (see Sec.~\ref{de}). 

One might be worried that the lepton and DM numbers will be immediately washed out since both $U(1)_L$ and $U(1)_\chi$ are explicitly broken by Yukawa interactions. However, the right-handed neutrinos and their partners reach chemical equilibrium long after the $SU(2)_\ell$ phase transition because of the smallness of their Yukawa couplings.  In the case of the SM neutrinos, this is implied by the small observed neutrino masses.  In the case of the neutrino partners,
this requires $\Gamma ( H \leftrightarrow \ell^\prime \nu' )$ be much less than the Hubble scale at the $SU(2)_\ell$ phase transition, 
which will be satisfied provided $y'_\nu \lesssim 10^{-6}$ for $u \sim$~TeV.
As a result both the lepton and DM number asymmetries survive until the electroweak transition. At that point the electroweak sphalerons see the effective lepton number deficit and transfer it into baryons through the Dirac leptogenesis mechanism \cite{Dick:1999je,Murayama:2002je}.

\subsection{Phase transition}

\subsubsection{Finite temperature effective potential}

The one-loop effective scalar potential at nonzero temperature can be written schematically in terms of the background field $u$ as
\bea
\hspace{-3mm}V(u, T) = V_{\rm tree}(u) + V_{{\rm 1\,loop}}(u, 0) + V_{\rm temp}(u, T) \ ,
\label{eff_pot}
\eea
where the first of the individual contributions is the tree-level part, 
\bea
V_{\rm tree}(u) = -\frac{1}{2} m^2 \,u^2 + \frac{1}{4} \lambda \, u^4 \ ,
\eea
where the mass parameter $m^2$ and quartic $\lambda$ schematically indicate combinations of those parameters from
the scalar potential, Eq.~(\ref{eq:V}).  The remaining terms on the right-hand side correspond to
the zero temperature Coleman-Weinberg correction and the one-loop finite temperature contribution. 

To calculate the Coleman-Weinberg term, 
we implement the cut-off regularization scheme and assume that the minimum of the one-loop potential and the $SU(2)_\ell$ Higgs mass are the same as their tree-level values (see, e.g. \cite{Quiros:1999jp}). The zero-temperature one-loop correction takes the form,
\bea
V_{1\,{\rm loop}}(u)  = \frac{1}{64\pi^2} \sum_i n_i \left\{m_i^4(u)\left[\log\left(\frac{m_i^2(u)}{m_i^2(v_\ell)}\right)-\frac{3}{2}\right]+2\,m_i^2(u)\,m_i^2(v_\ell)\right\} \ ,
\eea
where the sum is over all particles charged under $SU(2)_\ell$ and $n_i$ denoted the number of degrees of freedom, with an extra minus sign for the fermions.

Using the well-known formula for the one-loop finite temperature correction \cite{Quiros:1999jp},  the temperature-dependent piece is
\bea
V_{\rm temp}(u, T) = \frac{T^4}{4 \pi^2}\sum_i n_i (3\mp 1)\int_0^\infty d x \, x^2 \left[\log\left(1\mp e^{-\sqrt{x^2+ m_i^2(u)/T^2}}\right) - \log\left(1\mp e^{-x}\right)\right]\!.
\eea
In the expression above the sum is again over all fields, $n_i$ is the number of degrees of freedom including a factor of $-1$ for fermions; the minus signs are for bosons whereas the plus signs are for fermions.

\begin{figure}[t!]
\includegraphics[width=0.6\linewidth]{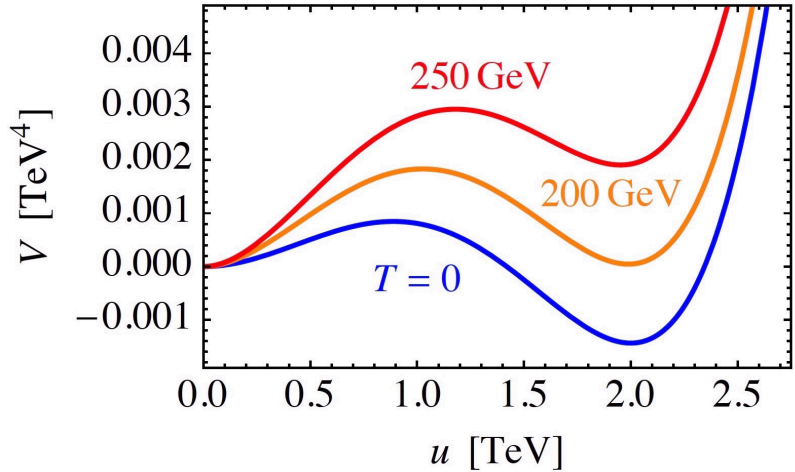} \vspace{-7mm}
\caption{
\small{Plot of the finite temperature effective potential $V(u, T_c)$ for $v_\ell = 2 \ \rm TeV$, $\lambda_{1} = 2\times 10^{-3}$, $g_\ell = 1$ and $T_c=200 \ \rm GeV$.}}
\label{fig:finite_temp}
\end{figure}

At this point, one can analyze the shape of the full effective potential for various temperatures. For successful baryogenesis the phase transition has to be 
strongly first order, $v_\ell (T_c) / T_c \gtrsim 1$, which favors small values of the effective quartic $\lambda$.
The constraint on $v_\ell$ coming from the LEP-II experiment is $v_\ell \gtrsim 1.7 \ \rm TeV$ \cite{Schwaller:2013hqa}. 
Finally, the critical temperature of the phase transition should not be lower than $\sim 175 \ {\rm GeV}$, in order to occur before the electroweak phase transition. 
Figure \ref{fig:finite_temp} shows the plot of $V(u, T_c)$ for sample parameter values fulfilling those constraints: $ g_\ell = 1, v_\ell = 2 \ \rm TeV$ and 
$\lambda = 2\times 10^{-3}$, leading to $T_c \sim 200 \ {\rm GeV}$.

\subsubsection{Bubble nucleation and diffusion equations}
\label{de}

A first order phase transition takes place at the critical temperature $T_c$. Bubbles of true vacuum are nucleated and then expand, eventually filling the entire universe. Following \cite{Shu:2006mm} and denoting the bubble radius by $R$ and the width by $L_w$, we examine an ansatz for the bubble profile,
\bea
u(r) &=& \frac{1}{2} u_c \left[1-\tanh\left(\frac{r - R}{L_w}\right)\right].
\eea
In this case the width of the bubble scales as $L_w \sim 1/T$. The expanding bubble is assumed to be large, so that to a good approximation we can analyze its evolution in one dimension, along the $z$ axis, taken to be perpendicular to the bubble wall. We place the bubble wall at $z=0$ with the broken phase on the  $z>0$ side. 
We adopt a bubble wall velocity of $v_w \approx 0.05\, c$.   

In the presence of CP violation, the $SU(2)_\ell$ instantons produce lepton and DM number asymmetries. In order to estimate their magnitude, a set of coupled diffusion equations is solved for particle number densities \cite{Cohen:1994ss,Shu:2006mm}. Since only leptons are affected by the presence of the new gauge group, there are 12 relevant equations in our case (see Appendix \ref{app:app_diff}) involving the following particle number densities,
\begin{align}
&{n(l)}= {n(e_L)} + {n(\nu}_L)\ , \ \ \ {n(e)}={n(e}_R)\ , \ \ \  {n(\nu)}={n(\nu}_R)\ , \ \ \ 
{n(\tilde{l})}=n(\tilde{e}_L) + n(\tilde{\nu}_L)\ , \ \ \ n(\tilde{e})=n(\tilde{e}_R)\ , \ \ \    n(\tilde{\nu})=n(\tilde{\nu}_R)\ , \nonumber\\
&{n(l}')=n({e}'_R) + {n(\nu}'_R)\ ,   \ \   {n(e')}={n(e}'_L)\ ,  \ \  {n(\nu')}={n(\nu}'_L)\ ,\ \ 
n(h)=n(h^+) + n(h^0)\ ,  \ \ n(\Phi^u)\!=n(\Phi_1^u)\ , \ \   n(\Phi^d)\! = n(\Phi^d_{2}) \ .
\end{align}
There are nine constraints on the particle number densities corresponding to the Yukawa equilibrium conditions, as well as four constraints coming from the instanton equilibrium requirement (see Appendix \ref{app:app_constraints}). However, not all of those constraints are independent. Only seven of the Yukawa equilibrium conditions and one of the instanton equilibrium conditions are linearly independent.

The diffusion equations contain diffusion constants for each particle species. For particles charged under the SM,
their magnitude has been estimated in Ref.~\cite{Joyce:1994zn}. Carrying out a similar calculation and taking the $SU(2)_\ell$ gauge coupling to be $g_\ell \approx 1$ we obtain the following estimates,
\bea
\hspace{-5mm}D_l = D_{\tilde{l}} \sim D_e = D_{\tilde{e}} \sim D_\nu = D_{\tilde{\nu}} \sim D_{\Phi^u} = D_{\Phi^d} \sim 25 / T\ , \ \ \ D_h = D_{l'} \sim 100 / T \ , \ \ \ D_{e'} \sim D_{\nu'} \sim 400 / T  \ .
\eea
Finally, we arrive at four coupled equations for the $l, \tilde{\nu}, \phi^u$ and $h$ particle number densities,
\bea
\label{dee1}
&& \hspace{-5mm}v_w\left[4n'(l)+4{n'(\tilde{\nu}})-2{n'({\Phi}^u})-n'(h)\right] - \frac{25}{T}\left[22\,n''(l) + 4\,n''(\tilde{\nu}) - 20 \,{n''(\Phi^u}) - n''(h)\right] = 0 \ , \\[5pt]
\label{dee2}
&& \hspace{-5mm}v_w\left[2n'(l)+{{n'(\Phi}^u})\right] - \frac{25}{T}\left[2\,n''(l) + {n''(\Phi^u}) \right] = \gamma_1 \, \theta(L_w-|z|) \ ,\\[5pt]
\label{dee3}
&& \hspace{-5mm}v_w\left[-n'(l)+6\,{n'(\tilde{\nu}})+{{n'(\Phi}^u})-\tfrac{3}{2}\,{n'(h})\right] - \frac{25}{T}\left[-n''(l) + 6\,n''(\tilde{\nu}) +{n''(\Phi^u}) -\tfrac{3}{2}\,n''(h)\right] = \gamma_2 \, \theta(L_w-|z|) \ ,\\[5pt]
&&  \hspace{-5mm}v_w\left[\tfrac{5}{2}\,{n'(h})\right] - \frac{25}{T}\left[13\,n''(h)\right] = 0 \ ,
\label{dee}
\eea
where the primes denote derivatives with respect to $z$, and $\gamma_1, \gamma_2$ are the CP-violating sources for the $SU(2)_\ell$ Higgs induced by the bubble wall.
The values for $\gamma_1$ and $\gamma_2$ in the two Higgs doublet model have been derived in \cite{Riotto:1995hh} and are given by 

\bea
\gamma_i(z) \approx \frac{\tilde{\lambda}_7}{32\pi}\Gamma_{\phi_i}T \frac{m^2_{12}}{m^3_{\phi_i}(T)}\partial_{t_z} \phi_i,
\eea
with $\tilde{\lambda}_7$ and $m^2_{12}$ free parameters from the scalar potential.
One can easily choose parameter values such that $\gamma_1 \approx \gamma_2 \approx 5 \times 10^{-5} \ {\rm GeV^4}$, 
which yields the observed ratio of the baryon to entropy ratio of $\sim 10^{-10}$, as discussed in the subsequent section.

\subsubsection{Lepton and dark matter asymmetries}

\begin{figure}[t!]
  \centering
      \includegraphics[width=0.6\textwidth]{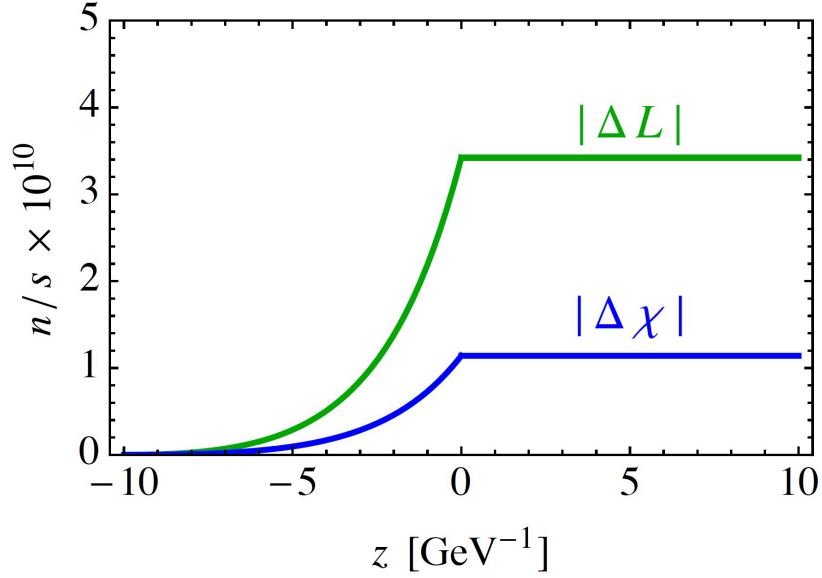}\vspace{-7mm}
  \caption{\small{SM lepton and DM particle number densities as a function of the spatial position $z$ assuming the bubble wall located at $z=0$.}}
  \label{fig:diff_sol}
\end{figure}

We are interested in the particle number densities corresponding to the conserved global $U(1)$ charges given in Table~{\ref{tab:global}}, i.e.,
\bea
 \Delta L (z) &=&  n(l) + n(\tilde{l}) + n(l') + n(e) + n(\tilde{e})  + n(e')\ ,\nonumber\\
\Delta \chi (z) &=& n(\nu) + n(\tilde{\nu}) + n(\nu')\ . 
\eea
Using the equilibrium conditions we arrive at,
\bea
 \Delta L (z) = 3\Big[n(l) + n(\tilde{\nu}) -\frac{1}{2}n(\Phi^u) - \frac{1}{2}n(h)\Big] \ ,
\eea
and
\bea
 \Delta \chi (z) = \Big[n(l) + n(\tilde{\nu}) - \frac{1}{2}n(\Phi^u) + \frac{1}{2}n(h)\Big] \ .
\eea
Figure \ref{fig:diff_sol}  shows the solution of the diffusion equations assuming $T_c = 200 \ \rm GeV$ and $\gamma_1 = \gamma_2 = 5 \times 10^{-5} \ {\rm GeV}^4$. The ratio of the produced lepton and DM asymmetries is,
\bea
\bigg|\frac{\Delta L}{\Delta \chi} \bigg| = 3 \ ,
\label{3/2}
\eea
roughly independent of the numerical values of $v_w$, $T_c$, $\gamma_1$, and $\gamma_2$.

\subsubsection{Baryon asymmetry}

The particle number densities in Fig.~\ref{fig:diff_sol}  are normalized to the entropy $s \approx (2\pi^2/45) g_*T^3$, with $g_*\sim 100$. For the above set of parameters, the ratio of the lepton number density and the entropy is roughly
$\Delta L/s\sim 3\times 10^{-10}$. The lepton asymmetry is
$n_L/s \approx 3\times 10^{-10}$.

The $SU(2)_\ell$ instantons shut off after the $SU(2)_\ell$ breaking concludes and the DM asymmetry freezes in. However, above the electroweak phase transition the SM sphalerons are active and they convert part of the SM lepton asymmetry to a baryon asymmetry. The baryon asymmetry generated by the sphalerons is \cite{Harvey:1990qw},
\bea
\Delta B = \frac{28}{79} \Delta L \ .
\label{B_asym}
\eea
This result only slightly depends on the lepton partner masses, with the effect minimized if those masses are below the electroweak scale. The final generated baryon asymmetry to entropy ratio is therefore,
\bea
\frac{n_B}{s} \approx 10^{-10} \ ,
\eea
in the correct ballpark to match observations.\\

\section{Dark matter}
\label{DMsec}

The DM is a mixed state, largely composed of the lightest $\tilde{\nu}_R$.  Through interactions with the SM Higgs, it picks up a
small component of electroweak doublet,
\bea
\chi_L=\nu'_L + \epsilon \,\tilde\nu_L \ , \nonumber\\
\chi_R=\tilde\nu_R + \epsilon \,\nu'_R \ ,
\label{chi}
\eea
with  $\epsilon \sim y_\nu v / (Y_\nu v_\ell) \ll 1$. 

In standard ADM models, the baryon and DM asymmetries are of similar size, depending on the exact form of the operators mediating them. 
Assuming the DM is relativistic at the decoupling temperature, this implies a DM candidate with a mass at  $\sim {\rm GeV}$ 
scale\footnote{We note that it is also possible to realize an ADM scenario with a DM mass of several TeV via Xogenesis \cite{Buckley:2010ui}. 
In that case the DM mass is approximately ten times the decoupling temperature and must be roughly $m_{Z'}/2$ to have a sufficiently large annihilation cross section.}. 
In particular, the  relation between the DM mass and the relic abundances is given by,
\bea
m_\chi = m_{p} \frac{\Omega_{\rm DM}}{\Omega_{\rm B}}  \bigg|\frac{\Delta B}{\Delta \chi}\bigg|\ .
\label{m_X}
\eea
From Eqs.~(\ref{3/2}) and (\ref{B_asym}) we obtain $|\Delta B / \Delta \chi| \approx 1$, which gives,
\bea
m_\chi \simeq 5 \ \rm GeV \ .
\eea

\begin{figure}[t!]
  \centering
      \includegraphics[width=0.6\textwidth]{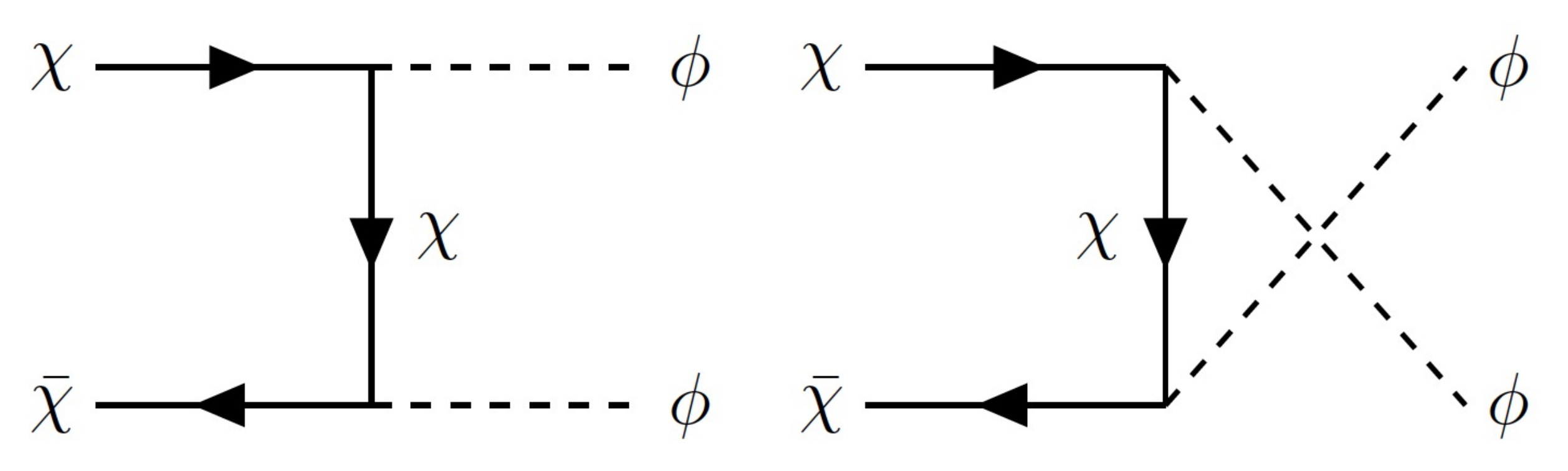}
  \caption{\small{Dark matter annihilation channels to the pseudoscalar component of $\Phi_2$.}}
  \label{fig:Higgs_ann}
\end{figure}

A mass of few GeV makes it difficult for the symmetric DM component to efficiently annihilate away, a generic challenge in ADM scenarios. 
We circumvent this issue by arranging for a light Higgs boson with $\sim$~GeV mass into which the DM can annihilate efficiently.
Provided the light scalar has a significant CP-odd component,
the Yukawa interactions in Eq.~(\ref{lagr}) and the DM content (\ref{chi}) imply that, to leading order in $\epsilon$, the coupling takes the form,
\bea
\mathcal{L}_{\rm DM} \approx Y_{\chi} \, \bar{\chi}\, \gamma^5 \phi \,\chi \ .
\eea
This provides a natural DM annihilation channel as shown in Fig.~\ref{fig:Higgs_ann}.
Using this interaction we obtain,
\bea
(\sigma v)_{\rm NR}  =  \frac{Y_\chi^4 \,m_\chi^6 v^2}{6\pi(2m_\chi^2-m_\phi^2)^4}\left(1-\frac{m_\phi^2}{m_\chi^2}\right)^{5/2}  .
\eea
Writing the thermally averaged annihilation cross section as
$
\langle\sigma_{\!A} v\rangle =    \sigma_0 \left({T}/{m_\chi}\right)
$
where
\bea
\sigma_0 =  \frac{Y_\chi^4 \,m_\chi^6}{2\pi(2m_\chi^2-m_\phi^2)^4}\left(1-\frac{m_\phi^2}{m_\chi^2}\right)^{5/2} \ ,
\eea
the present energy density of the symmetric component of the DM particles is \cite{Kolb:1990vq,Griest:1990kh},
\bea
\Omega_{\chi} h^2 \simeq\left( \frac{1.75\times 10^{-10}}{{\rm GeV^2}}\right) \frac{1}{\sigma_0  \sqrt{g_*}} \left(\frac{m_\chi}{T_f} \right)^2,
\label{ann2}
\eea
where $T_f$ is the freeze-out temperature and $g_*$ is the number of relativistic degrees of freedom. 
For $m_\phi \approx 1 {\rm \ GeV}$, the remnant symmetric component will be subdominant to the asymmetric component produced by the
$SU(2)_\ell$ phase transition provided the Yukawa coupling satisfies,
\bea
{Y}_{\nu}\gtrsim 0.1  \ .
\label{relationg}
\eea

Light scalar bosons of mass $\sim$~GeV are somewhat unexpected from a potential whose overall energy scale is characterized by $v_\ell \sim$~TeV.
However, they can be realized provided the quartics are all small (which is also favored by the need for a strongly first order phase transition) and/or
$\tan \beta$ is large, indicating that $v_1 \gg v_2$.  While a detailed analysis of the scalar sector is beyond the scope of this work, it is a generic prediction
that there will be light ($\sim$ GeV) scalar particles with weak ($\sim 10^{-3}$) couplings to leptons.  Such particles are typically not currently constrained by
low energy experiments, but may be accessible in the future \cite{Krnjaic:2015mbs}.

The coupling of $\phi$ to the SM Higgs via $\lambda |\Phi|^2 |H|^2$ terms in the scalar potential allows for thermal equilibrium between $\phi$ and the SM provided $\Gamma(\phi \rightarrow \mu \bar{\mu})$ is greater than the expansion rate at temperatures $T\sim 1 ~ \rm GeV$.  This holds true for $\lambda \gtrsim 10^{-6}$.  While the level of scalar mixing corresponding to the  lower limit is too small to be observable at the LHC, larger values could be detectable. 

\begin{figure}[t!]
  \centering
      \includegraphics[width=0.5\textwidth]{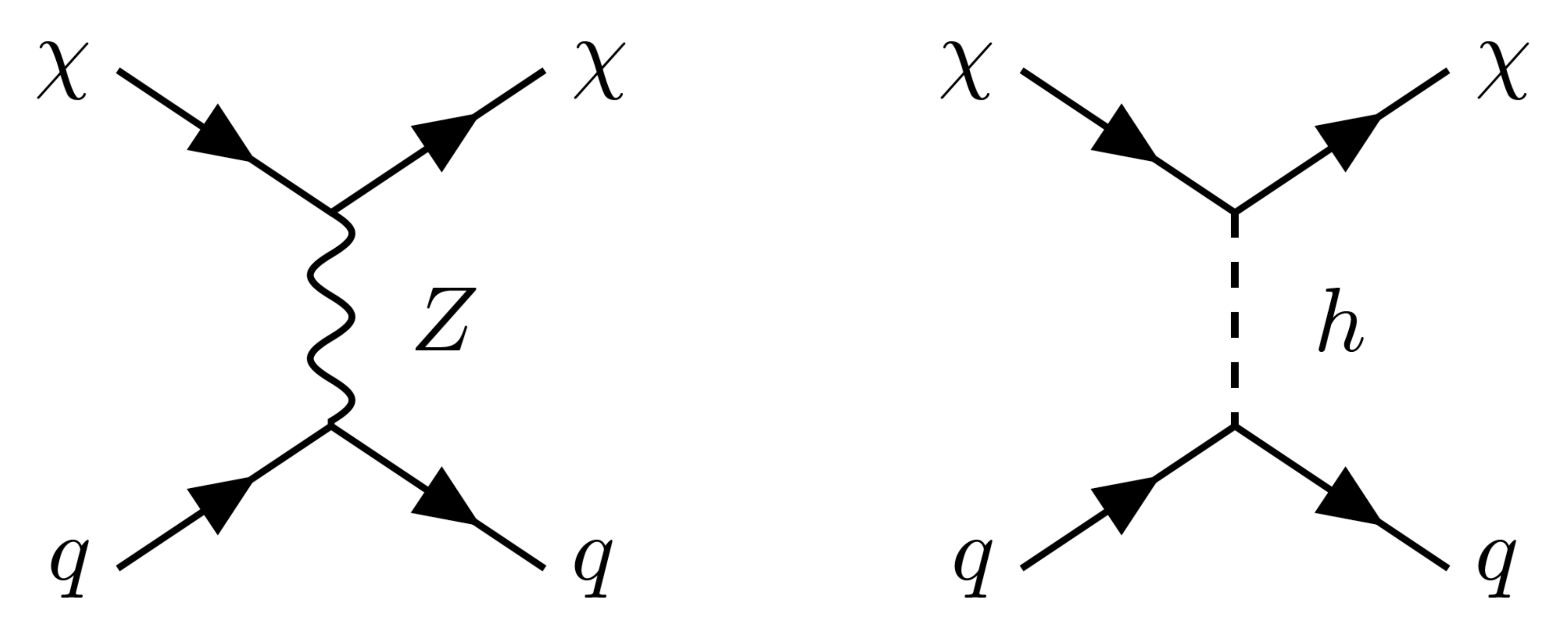}
\caption{\small{Diagrams contributing  to  DM interacting with quarks.}}
  \label{fig:111}
\end{figure}

The primary mediator for dark matter scattering with heavy nuclei is the
$Z'$ gauge boson, which does not interact with quarks and therefore does not appear in tree-level DM direct detection diagrams 
(though contributions do appear at one loop). As a result, the corresponding bound on $v_\ell$ set by the null search results from the CDMSlite 
experiment \cite{Agnese:2015nto} is much less stringent than the collider constraint from LEP-II of $v_\ell \gtrsim 1.7 \ \rm TeV$. 
The DM direct detection diagrams involving the SM gauge bosons are shown in Fig.~\ref{fig:111}. The calculation of the spin-independent direct detection cross section closely follows the results of \cite{Fornal:2016boa} for the electroweak diagrams, which, combined with the CDMSlite bounds, 
provides an upper limit on the doublet admixture parameter $\epsilon \lesssim 0.3$, consistent with the assumption that $y'_{\nu,e} v \ll Y_{\ell, \nu, e} v_\ell$.

\section{Conclusions}

We have investigated a novel extension of the Standard Model, in which $SU(3) \times SU(2) \times U(1)$ is supplemented 
by a non-Abelian gauge group $SU(2)_\ell$, in which the leptons are promoted to doublets by introducing lepton partner fermions. 
The doublet partners pair up with extra $SU(2)_\ell$ singlet fields needed to cancel the anomalies and develop vector-like masses after $SU(2)_\ell$ breaking.
We find that the global symmetries produce a 
viable dark matter candidate, which is the lightest of the lepton partners and stable due to a residual $U(1)_\chi$ global symmetry. 
In addition, it can explain baryogenesis through the breakdown of the new gauge group and naturally fits into an asymmetric dark matter framework. 

We explicitly analyze the details of the dynamics of $SU(2)_\ell$ phase transition, during which nonperturbative interactions mediated by $SU(2)_\ell$ instantons
violate both lepton and dark matter numbers, leading to a correlated asymmetry in both sectors.  We find that the wide (and relatively unconnected to
electroweak observables) parameter space easily allows for a first order phase transition with sufficient asymmetry generated in the leptons to
lead (through the electroweak sphalerons) to a baryon asymmetry in line with observation.  Provided the mass of the dark matter is a few GeV, its correlated
abundance will also match cosmological observations.

Both the lepton partners and the additional neutral gauge bosons
are colorless, and thus are not efficiently produced at the LHC.
The most stringent constraints are bounds on the $Z'$, which contributes to $e^+ e^-$ at LEP-II.
It would be interesting to see how a future high energy $e^+ e^-$ collider could shed light on such a scenario, and whether it could preclude enough of
parameter space to say something definitive about its potential realization as the mechanism for baryogenesis. 

It is also of interest to explore similar extensions
of the Standard Model based on generalizations of baryon number.  This is currently work in progress \cite{West:2017boa}.

\subsection*{Acknowledgments}

This research was supported in part by the NSF grant PHY-1316792. BF also acknowledges partial support from the DOE grant DE-SC0009919 
and is grateful to the Institute of Advanced Studies at the Nanyang Technological University in Singapore for hospitality during the completion of this work.
TMPT is grateful for helpful conversations with M. Buckley, Y. Kahn, G. Krnjaic, M. Perelstein, M. Peskin, and C.E.M. Wagner.

\appendix

\section{Diffusion equations}

\label{app:app_diff}

The rates at which the particle densities change are described by 12 diffusion equations. We denote the common rates for the
$SU(2)_\ell$ instanton induced interactions by $\Gamma$.
Neglecting the neutrino and neutrino partner Yukawas, the equations for the particle number densities $n(l), n(\tilde{l}), n(l')$ are given in simplified notation (all particle number densities labeled by their species, e.g. $\tilde{\ell}\equiv n(\tilde{\ell})$) by,
\bea
&&\hspace{-5mm}\dot{l}-  D_l \nabla^2 l = - \Gamma  \left[4l+2\tilde{l}-4({\nu} + {e})-2(\tilde{\nu} + \tilde{e})\right]  -  \Gamma_{Y_l} \left(\frac{l}{2}-\frac{\Phi^u}{2}-\frac{l'}{2}\right) -  \Gamma_{y_e} \left(\frac{l}{2}-\frac{h}{4}-e\right)  ,\\[6pt]
&&\hspace{-5mm}\dot{\tilde{l}}-  D_{\tilde{l}} \nabla^2 \tilde{l} = - \Gamma  \left[4\tilde{l}+2{l}-4(\tilde{\nu} + \tilde{e})-2({\nu} + {e})\right] -  \Gamma_{Y_l} \left(\frac{\tilde{l}}{2}-\frac{\Phi^d}{2}-\frac{l'}{2}\right) -  \Gamma_{y_{e}} \left(\frac{\tilde{l}}{2}-\frac{h}{4}-\tilde{e}\right),\\
&&\hspace{-5mm}\dot{l}'-  D_{l'} \nabla^2 l' = - \Gamma_{y'_e} \left(\frac{{l'}}{2}-\frac{h}{4}-e'\right)+  \Gamma_{Y_l} \left(\frac{l}{2}-\frac{\Phi^u}{2}-\frac{l'}{2}\right) +  \Gamma_{Y_l} \left(\frac{\tilde{l}}{2}-\frac{\Phi^d}{2}-\frac{l'}{2}\right).
\eea
For $\nu, \tilde{\nu}, \nu'$,
\bea
&&\hspace{-5mm}\dot{\nu} -  D_{\nu} \nabla^2 \nu=  - \ \Gamma_{Y_\nu} \left({\nu}-\frac{\Phi^u}{2}-{\nu'}\right)-  \Gamma  \left(3{\tilde{\nu}}+\tilde{e}+2{{e}}-{l}-2{\tilde{l}}\right), \ \ \ \ \ \ \ \ \ \\ 
&&\hspace{-5mm}\dot{\tilde{\nu}} -  D_{\tilde{\nu}} \nabla^2 \tilde{\nu} =  - \ \Gamma_{Y_\nu} \left({\tilde{\nu}}-\frac{\Phi^d}{2}-{\nu'}\right)-  \Gamma  \left(3{\nu}+{e}+2\tilde{e}-\tilde{l}-2{l}\right), \\ 
&&\hspace{-5mm}\dot{\nu}' -  D_{\nu'} \nabla^2 \nu' =   \Gamma_{Y_\nu} \left({\nu}-\frac{\Phi^u}{2}-{\nu'}\right) +  \Gamma_{Y_\nu} \left({\tilde\nu}-\frac{\Phi^d}{2}-{\nu'}\right).  
\eea
For $e, \tilde{e}, e'$,
\bea
&&\hspace{-5mm}\dot{e} -  D_e \nabla^2 e =  - \ \Gamma_{Y_e} \left({e}-\frac{\Phi^u}{2}-{e'}\right)   +  \Gamma_{y_e} \left(\frac{l}{2}-\frac{h}{4}-e\right)-  \Gamma  \left(3\tilde{e}+\tilde{\nu}+2{{\nu}}-{l}-2\tilde{l}\right), \ \  \\  
&&\hspace{-5mm}\dot{\tilde{e}} -  D_{\tilde{e}} \nabla^2 \tilde{e} =  - \ \Gamma_{Y_e} \left({\tilde{e}}-\frac{\Phi^d}{2}-{e'}\right)  +  \Gamma_{y_{e}} \left(\frac{\tilde{l}}{2}-\frac{h}{4}-\tilde{e}\right)-  \Gamma  \left(3{{e}}+{{\nu}}+2\tilde{\nu}-{\tilde{l}}-2{{l}}\right), \\ 
&&\hspace{-5mm}\dot{e}' -  D_{e'} \nabla^2 e' =  \Gamma_{y'_e} \left(\frac{{l'}}{2}-\frac{h}{4}-e'\right)  +  \Gamma_{Y_e} \left({e}-\frac{\Phi^u}{2}-{e'}\right) + \Gamma_{Y_e} \left({\tilde e}-\frac{\Phi^d}{2}-{e'}\right).
\eea
Finally, for the particle number densities $\Phi^u, \Phi^d$ and  $h$,
\bea
&&\hspace{-5mm}\dot{\Phi}^u -  D_{\Phi^u} \nabla^2\Phi^u =   \gamma_1 + \Gamma_{Y_l} \left(\frac{l}{2}-\frac{\Phi^u}{2}-\frac{l'}{2}\right)+ \Gamma_{Y_\nu} \left({\nu}-\frac{\Phi^u}{2}-{\nu'}\right)  + \Gamma_{Y_e} \left({e}-\frac{\Phi^u}{2}-{e'}\right), \ \ \ \ \ \ \ \\
&&\hspace{-5mm}\dot{\Phi}^d -  D_{\Phi^d} \nabla^2\Phi^d =  \gamma_2 + \Gamma_{Y_l} \left(\frac{\tilde{l}}{2}-\frac{\Phi^d}{2}-\frac{l'}{2}\right) +  \Gamma_{Y_\nu} \left({\tilde{\nu}}-\frac{\Phi^d}{2}-{\nu'}\right)+ \Gamma_{Y_e} \left({\tilde{e}}-\frac{\Phi^d}{2}-{e'}\right),  \\
&&\hspace{-5mm}\dot{h} -  D_h \nabla^2 h = \Gamma_{y_e} \left(\frac{l}{2}-\frac{h}{4}-e\right) +  \Gamma_{y_e} \left(\frac{\tilde{l}}{2}-\frac{h}{4}-\tilde{e}\right)  - \Gamma_{y'_e} \left(\frac{{l'}}{2}-\frac{h}{4}-e'\right).
\eea
where $\gamma_1$ and $\gamma_2$ are the CP-violating sources.

\section{Constraints on particle number densities}

\label{app:app_constraints}

The equilibrium conditions emerging from the Yukawa terms are,
\bea
&&\frac{l}{2} - \frac{\Phi^u}{2} - \frac{l'}{2} = 0 \ , \ \ \ \ \ \ \frac{\tilde{l}}{2} - \frac{\Phi^d}{2} - \frac{l'}{2} = 0  \ , \ \ \ \ \ \   \nu - \frac{\Phi^u}{2} - \nu' = 0 \ , \ \ \ \ \ \ \  \tilde{\nu} - \frac{\Phi^d}{2} - \nu' = 0 \ , \nonumber\\
&& e - \frac{\Phi^u}{2} - e' = 0 \ , \ \ \ \ \ \ \  \ \tilde{e} - \frac{\Phi^d}{2} - e' = 0 \ , \ \ \ \ \ \  \frac{l}{2} - \frac{h}{4} - e = 0 \ , \ \ \ \ \ \ \frac{\tilde{l}}{2} - \frac{h}{4} - \tilde{e} = 0  \ , \ \ \ \ \ \  \frac{l'}{2} - \frac{h}{4} - e' = 0 \ .
\eea
And those from the instanton-induced interactions can be written as,
\bea
&&l - \nu - e = 0 \ , \ \ \ \ \ \ \ \frac{l}{2} +\frac{\tilde{l}}{2} - \nu - \tilde{e} = 0 \  , \ \ \ \ \ \ \  \tilde{l} - \tilde{\nu} - \tilde{e} = 0 \ , \ \ \ \ \ \ \ \frac{l}{2} + \frac{\tilde{l}}{2} - \tilde{\nu} - {e} = 0 \ .
\eea
As discussed in the main text, two of the Yukawa equilibrium conditions and three of the instanton conditions are linearly dependent on the others. 
Therefore, there are only four independent particle number densities. For example one can choose them to be $l$, $\tilde{\nu}$, $h$ and $\Phi^u$, in which case the other particle densities are given by,
\bea
&& \ \ \ \ \ \ \tilde{l}= 2\tilde{\nu} - \frac{h}{2} \ , \ \ \ \ \, \  l'= l - \Phi^u \ , \ \ \  \ \   \nu = \frac{l}{2} + \frac{h}{4} \ ,  \ \ \ \ \ \ \nu'= \frac{l}{2} - \frac{\Phi^u}{2} + \frac{h}{4} \ , \nonumber\\
&&\ \ \ \ \ \  e = \frac{l}{2}  - \frac{h}{4} \ , \ \ \ \ \ \  \tilde{e}= \tilde{\nu}- \frac{h}{2} \ ,  \ \ \ \ \ \ e' = \frac{l}{2} - \frac{\Phi^u}{2} - \frac{h}{4} \ , \ \ \ \   \ \Phi^d = 2\tilde{\nu} - l + \Phi^u - \frac{h}{2} \ .
\eea

\bibliography{SU2lepton}

\end{document}